# Solving bandwidth-guaranteed routing problem using routing data

Cao Thai Phuong Thanh[1], Ha Hai Nam[2] and Tran Cong Hung[2]

[1]Saigon University, Vietnam

[2] Post & Telecommunications Institute of Technology, Vietnam

## Abstract

*This paper introduces a traffic engineering routing algorithm that aims to accept as many routing demands as possible on the condition that a certain amount of bandwidth resource is reserved for each accepted demand. The novel idea is to select routes based on not only network states but also information derived from routing data such as probabilities of the ingress egress pairs and usage frequencies of the links. Experiments with respect to acceptance ratio and computation time have been conducted against various test sets. Results indicate that the proposed algorithm has better performance than the existing popular algorithms including Minimum Interference Routing Algorithm (MIRA) and Random Race based Algorithm for Traffic Engineering (RRATE).*

## 1. Introduction

There are increasingly more network applications such as live stream and online games that require certain quality of service (QoS) guarantees. Apart from meeting the QoS requirements, network providers need to optimize their network performance in order to effectively fulfil as many customer demands as possible. Therefore, traffic engineering (TE), which manages network activities with dual objectives of QoS satisfaction and network optimization, becomes important. Routing is a powerful technique of TE as it allows for controlling network data flows.

In general, routing algorithms are categorized into proactive and reactive ones. The former pre-selects paths for routing demands based on fixed network information, while the later uses dynamic network states to establish routes upon receiving demands. As a result, reactive algorithms adapt to traffic demands and have good routing performance [1]. Furthermore, although there are several types of QoS criteria such as bandwidth, delay, and loss ratio, most research considers bandwidth as a primary constraint as the others can be efficiently converted into bandwidth demand [2]. Therefore, this paper focuses on reactive bandwidth-guaranteed routing problem. Besides network states (e.g. network topology, link residual bandwidth), which are used in existing solutions [3], the proposed algorithm introduces a novel idea to exploit historical routing data such as request probability and link usage frequency for route selection.

The rest of the paper is organized as follows. Section 2 presents the definition of the TE routing problem and a review of related work. The proposed algorithm named Traffic Engineering routing Algorithm with Routing Data (TEARD) is discussed in section 3. Section 4 describes the experimental setup and compares the proposed algorithm with some popular ones with respect to acceptance ratio of demands and average computation time. Finally, section 5 discusses conclusions and future work.

## 2. PROBLEM DEFINITION & RELATED WORK

### 2.1. Problem Definition

A network topology with *n* nodes and *m* links is considered. Each link has its own capacity and residual bandwidth at a given time. A routing demand, which requires a path with certain bandwidth from an ingress node to an egress node, is handled by the routing algorithm. The algorithm sequentially processes demands with an assumption that network states such as topology, link bandwidths and ingress-egress (*ie*) pairs are available. However, routing demands are not known in prior. Table 1 lists the mathematical notations used to describe the problem.

Table 1. Problem notations

| Symbol | Description |
|--------|-------------|
| $G(N, L)$ | A direct graph presents the network topology |
| | $N$ ($|N| = n$) is a set of nodes |
| | $L$ ($|L| = m$) is a set of links |
| $IE$ | A set of all ingress egress pairs |
| $c(l)$ | Capacity bandwidth of link $l$ |
| $r(l)$ | Residual bandwidth of link $l$ |
| $d(i, e, b)$ | A traffic demand requesting $b$ bandwidth units from an ingress node $i$ to an egress node $e$ |
| $p_{ie}$ | A routing path from $i$ to $e$ |
| $P_{ie}$ | A set of all paths from $i$ to $e$ |

The objective of TE routing algorithm is to route as many demands as possible on the condition that each established route will reserve an amount of bandwidth resource for a period of time (i.e. bandwidth for each route is guaranteed). Since ingress-egress pairs have commodity integral flows, the TE routing problem is NP-hard [4]. Most of reactive routing algorithms first calculate link weights based on network states then use shortest path algorithms (e.g. Dijkstra or Bellman-Ford) to select the least weighted route. Table 2 generalizes the steps of reactive routing algorithms.

Table 2. General steps of TE routing algorithms

| Input | A network graph $G(N, L)$ with necessary information e.g. link bandwidths. A traffic demand $d(i, e, b)$ |
|-------|-------------|
| Output | A satisfied bandwidth path from $i$ to $e$, $p_{ie}$, toward the optimal goal of maximizing the number of accepted demands. Or no route satisfying the demand |
| Steps | 1. Calculate link weights $w(l)$ 2. Temporarily remove links that have residual bandwidth less than $b$. 3. Find the least weight path $p_{ie}$. If found then return $p_{ie}$, otherwise rejects the demand. |

## 2.2. Related Work

The simplest and most widely used routing solution is Minimum Hop Algorithm (MHA) [5]. As its name implies, MHA selects paths that have minimum hop counts for routing. It means the same shortest path is chosen for each ingress egress pair until at least one of its link cannot satisfy bandwidth demands. This static selection scheme results in network bottleneck and under-utilization.

Minimum Interference Routing Algorithm (MIRA) \cite{MIRA_2000} exploited knowledge of ingress egress pairs so that routing paths of one pair interfere as little as possible with paths of the others. The interference is measured based on the maxflow-mincut theory [6]. Particularly, when a routing demand arrives, MIRA identifies mincut sets for all of the ingress egress pairs except the one being requested. Links that belong to a mincut set are considered critical because if they are used to route data then maxflow of the corresponding pair will decrease. It means critical links will interfere with ingress egress pairs. Therefore, MIRA sets link weights as the number of times the links are critical (equation 1). This idea allows MIRA to accept noticeably more requests than MHA. However, computation time of MIRA is also significantly longer than MHA's due to the maxflow-mincut calculation.

$$w_{ie}(l) = \sum_{(s,d) \in IE \setminus (i,e)} \alpha_{sd} \quad \text{if } l \text{ is critical of } (s,d) \tag{1}$$

$\alpha_{sd}$ is a parameter reflecting the importance of $(s,d)$

Random-based Routing Algorithm for Traffic Engineering (RRATE) [7] employed a machine learning technique - random race - to improve the MIRA computation time. RRATE has two stages for each ingress egress pair: learning and post-learning. In the learning stage, upon receiving a routing demand, costs of $k$ pre-selected paths are calculated based on maxflow-mincut criticality (the same as MIRA) and residual bandwidths (equation 2). The least cost path is then selected and its racing reward is accumulated. The race between those $k$ paths in term of reward values will end when there is a path whose reward reaches a pre-defined value $N$. After that, the corresponding pair moves into the post-learning stage and costs are not calculated any more. Alternatively, demands of that pair will be routed by the path having maximum racing value and its links satisfy bandwidth constraint. This post-learning stage reduces the computing time of RRATE compared to MIRA.

$$cost(p_{i\_ie}) = k_1.C_{i\_ie} + k_2/R_{i\_ie} \tag{2}$$

$$C_{i\_ie} = \sum_{l \in p_{i\_ie}} \sum_{(i,e) \in IE} v_l$$

$$v_l = \begin{cases} 1 & \text{if } l \text{ is critical of } (i,e) \\ 0 & \text{if } l \text{ is not critical of } (i,e) \end{cases}$$

$$R_{i\_ie} = max_{l \in p_{i\_ie}}(r(l) - b)$$

$k_1, k_2$ are the moderation parameters

Additionally, authors of [8] proposed another two-phase routing algorithm called Bandwidth Guaranteed MPLS Routing Algorithm (BGMRA). In BGMRA, link criticality is considered as occurrences of links in all of possible paths. It means the more paths a link belongs to, the more critical it is (i.e. the more interference it causes). Because BGMRA uses only network topology to calculate criticality, this process is done independently of demand arrival and called offline phase. On the other hand, the online phase is the process to select route for arriving demands. Equation 3 shows the calculation of link weights in the BGMRA online phase. BGMRA has low computation time because link criticality is calculated in the offline phase and recalculation is needed only when the network topology changes.

$$criticality(l) = \sum_{(i,e) \in IE} \sum_{p_{ie} \in P_{ie}} \frac{v_l}{|P_{ie}|} \qquad (3)$$

$$v_l = \begin{cases} 0 & \text{if } l \notin p_{ie} \\ 1 & \text{if } l \in p_{ie} \end{cases}$$

$|P_{ie}|$ is the number of all paths from i to e

$$w(l) = criticality(l).\frac{1}{r(l)}$$

## 3. PROPOSED ALGORITHM

This section describes the proposed algorithm named Traffic Engineering routing Algorithm with Routing Data (TEARD). In order to make the algorithm adaptive to routing demands, link weights are calculated from not only network topology and residual bandwidths but also from routing data.

Firstly, ingress egress pairs are considered. For each ie pair, link criticality is determined as the occurrence rate of the link in all paths of the pair. For instance, if a pair *ie* has 5 paths and a link *l* appears in 3 of them, then the criticality of link *l* for the pair *ie* is *crit$_{ie}$(l)=3/5*. This criticality is calculated in the offline phase because it is determined by the network topology only. Furthermore, in the online phase, the link criticality for each pair is multiplied by probability of the pair being requested. This modification makes the critical values dynamically adapt to actual routing requests. For example, when link *l* appears in many paths of pair *ie*, the critical value *crit$_{ie}$(l)* is high and the algorithm will try to avoid this link. However, if *ie* is infrequently requested, such avoidance may have negative effect on the overall routing performance. In this case, the multiplication by low probability of *ie* helps reducing *crit$_{ie}$(l)*. The link criticality for *ie* pairs is calculated as follows:

$$crit_1(l) = \sum_{ie \in IE} (crit_{ie}(l) * prob(ie) * 100) \qquad (4)$$

Where *crit$_{ie}$(l)* is the criticality of link *l* for the ingress egress pair *ie*.

$$crit_{ie}(l) = \frac{\sum_{p_{ie} \in P_{ie}} v_l}{|P_{ie}|} \qquad (5)$$

$$v_l = \begin{cases} 0 & \text{if } l \notin p_{ie} \\ 1 & \text{if } l \in p_{ie} \end{cases}$$

$|P_{ie}|$ is the number of all paths from i to e

*prob(ie)* is the probability of the pair $ie$ being requested. In this paper, *prob(ie)* is statistically calculated from *ie* pairs of historical routing requests.

Secondly, bandwidth information, especially residual bandwidth, plays an important role in route selection. In order to prevent bottleneck, the less residual bandwidth a link has, the less likely it should be chosen for new paths i.e. its critical value should be high. Similarly, a link that has used large of its bandwidth also should not been chosen. As a result, dynamic bandwidth states are taken into link criticality as follow:

$$crit_2(l) = \frac{used(l)}{r(l)} = \frac{c(l) - r(l)}{r(l)} * 100 \tag{6}$$

Thirdly, usage frequencies of links within the routing process also contribute to critical value as shown in equation 7. The idea is to balance link selection by setting higher value to a link that has been selected more times.

$$crit_3(l) = \frac{|P_l|}{|P|} * 100 \tag{7}$$

Where $|P|$ is the total number of established paths and $|P_l|$ is the number of established paths that contain link $l$.

Finally, equation 8 combines three above critical parts into the TE weight of link $l$; and table 3 describes steps of the proposed algorithm TEARD.

$$w(l) = k_1.crit_1(l) + k_2.crit_2(l) + k_3.crit_3(l) \tag{8}$$

$k_1, k_2, k_3$ are moderation parameters

$0 < k_1, k_2, k_3 < 1$ and $k_1 + k_2 + k_3 = 1$

Table 3. The Traffic Engineering Routing Algorithm using Routing Data (TEARD)

| Input | A network graph $G(N, L)$ |
| | A traffic demand $d(i, e, b)$ |
| Output | A satisfied bandwidth path from $i$ to $e$, $p_{ie}$ |
| | Or no route. |
| Offline phase | 1. For each pair $ie$, calculate $crit_{ie}(l)$ of every link using equation 5 |
| Online phase | For each $d(i, e, b)$: |
| | 1. Calculate link weights $w(l)$ using equation 8 |
| | 2. Temporarily remove links whose $r(l) < b$ |
| | 3. Apply Dijkstra algorithm to find the least weighted path from $i$ to $e$ |
| | If found: return the path $p_{ie}$ and update necessary information. |
| | Else: reject the demand (no route). |

# 4. PERFORMANCE EVALUATION

The proposed algorithm TEARD is evaluated against three popular TE routing algorithms: MHA, MIRA, and RRATE. Two performance metrics are considered: acceptance ratio and average computation time. The acceptance ratio reflects the percentage of demands for which algorithms are able to establish bandwidth-reserved path. According to the problem definition, routing algorithms aim to maximize this metric. Meanwhile, the computation time is the average time needed for algorithms to handle routing demands. The average time is computed in the online phase and should be minimized.

## 4.1. Experimental Setup

Algorithms are evaluated in two routing scenarios: static and dynamic ones. In static scenario, routing demands arrive at the same rate. If the arrived demands are accepted they will statically hold bandwidth resource along their paths. On the other hand, paths in dynamic scenario will release bandwidths after a certain holding time. Arrival time of dynamic demands is randomly distributed according to the Poisson process with mean requests per time unit and their holding time follows the Exponential distribution with mean $\mu$ time units. There are respectively 1000 and 2000 demands for static and dynamic scenarios.

Figure 1 shows three experimental network topologies: MIRA which is adapted from previous TE routing work e.g. [2], [8]; CESNET which is inherited from a real MPLS topology [9], and ANSNET - an topology based on United State geography [10]. All topologies contain bidirectional links. Capacities, ingress egress pairs, and demanded bandwidths for each topology are shown below:

- MIRA (fig. 1(a)) has 4800 and 1200 bandwidth units respectively for thicker and thinner links. Ingress egress pairs are (0, 12), (4, 8), (3, 1), (4, 14) and bandwidths are randomly demanded between 5, 11, 17, and 23 units.
- CESNET (fig. 1(b)) has 10000 (thicker links) and 1000 (thinner links) bandwidth units. Ingress egress pairs are (0, 18), (1, 11), (3, 16), (4, 7), (5, 13), (6, 19), (15, 0), (19, 8) and demanded bandwidths are 40, 80, 120, 160 units.
- All links of ANSNET (fig. 1(c)) have the same capacity of 2000 bandwidth units. There are 10 ingress egress pairs: (0, 28), (1, 13), (2, 30), (3, 22), (4, 10), (6, 30), (8,23), (21, 5), (17, 5), and (20, 16). Meanwhile demanded bandwidths are 20, 30, 40, and 50 units.

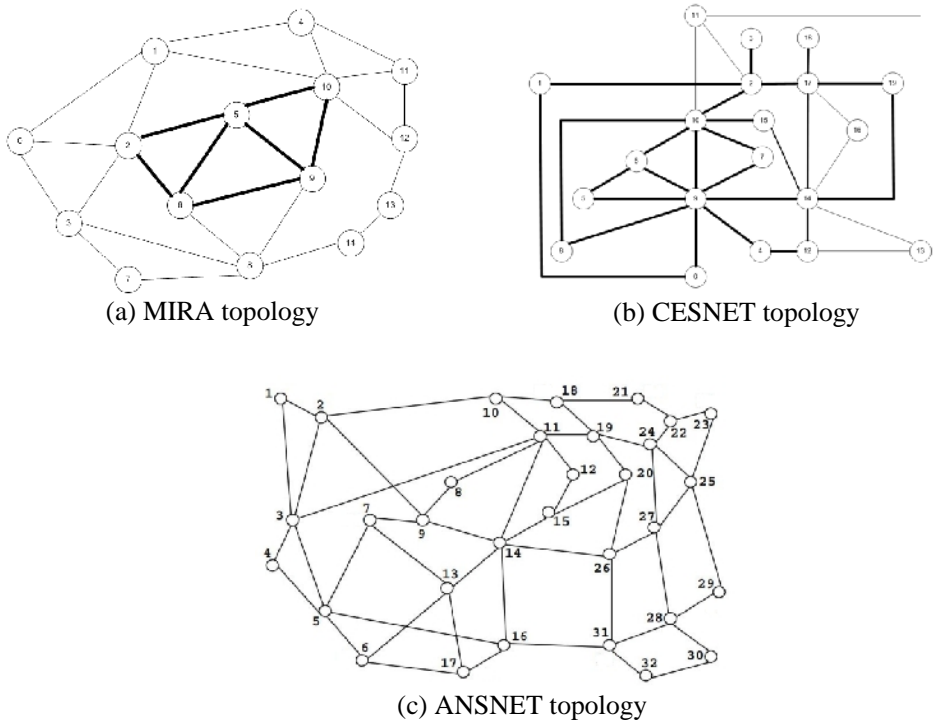

(a) MIRA topology          (b) CESNET topology

(c) ANSNET topology

Figure 1.  Simulated network topologies

Furthermore, RRATE takes moderation parameters $k_1=k_2=0.5$, number of pre-selected path $k=30$, and racing threshold $N=15$. Those parameter values are chosen for they give the best results after trying different values. Meanwhile, TEARD takes moderation parameters $k_1=0.3$, $k_2=0.4$, $k_3=0.3$. Values of TEARD parameters are discussed in later subsection.

**4.2. Performance Results**

Firstly, experimental results on MIRA topology are presented. Four ingress egress pairs of MIRA are demanded by the approximate rates of 10%, 20%, 30%, and 40%. Dynamic demands have the Poisson arrival time $=40$ and the Exponential holding time $\mu = 20$. Table 4 shows numeric results of one static and one dynamic tests.

Table 4. Results of acceptance ratio (in percentage) - computation time (in milliseconds) subject to number of demands (NoD) on MIRA topology

(a) Static scenario

| NoD | MHA | MIRA | RRATE | TEARD |
|---|---|---|---|---|
| **100** | 100- 0.19 | 100- 2.23 | 100- 3.05 | 100- 0.29 |
| **500** | 94.40- 0.10 | 94.40- 2.16 | 94.40- 1.91 | 94.60- 0.24 |
| **1000** | 56.90- 0.04 | 58.70- 1.02 | 59.30-0.31 | 61.20- 0.11 |

(b) Dynamic scenario

| NoD | MHA | MIRA | RRATE | TEARD |
|---|---|---|---|---|
| **100** | 100- 0.22 | 100- 2.36 | 100- 3.10 | 100- 0.30 |
| **1000** | 71.30- 0.09 | 72.80- 2.22 | 74.60- 0.62 | 74.30- 0.23 |
| **2000** | 63.60- 0.08 | 64.70- 2.40 | 64.85- 0.34 | 67.10 - 0.23 |

Figure 2 plots acceptance ratios versus last 300 demands as well as average computing times of static scenario on MIRA topology. It is clearly observable that TEARD accepts more demands than the other algorithms. Particularly, after 1000 demands, acceptance ratio of TEARD is the highest (61.2%). Next is RRATE (59.3%) then MIRA (58.7%), and MHA has the lowest acceptance at 56.9%. However, MHA is the fastest algorithm, meanwhile MIRA is the lowest due to its maxflow-mincut calculation. For instance, in figure 2(b) computing time of MHA is faster than MIRA about 25 times. Figure 2(b) also depicts that RRATE requires long time to handle first several hundreds of demands, but after that learning stage the time of RRATE considerably decreases. Nevertheless, TEARD achieves good computation time (average of 0.1 ms) that is comparable to MHA (0.04 ms) and much lower than both MIRA (1.0 ms) and RRATE (0.3 ms).

Moving to the second topology CESNET, the probabilities of eight ingress egress pairs are 5%, 10%, 15%, and 20% (duplicated). Dynamic demands have $=80$ and $\mu = 30$. Those dynamic time parameters are selected based on topology sizes where the bigger network has more demands and longer holding time. Table 5 shows examples of results on CESNET topology.

Table 5.  Experiment results on CESNET topology

(a) Static scenario

| NoD | MHA | MIRA | RRATE | TEARD |
|---|---|---|---|---|
| **100** | 100- 0.20 | 100- 8.36 | 100- 10.42 | 100- 0.54 |
| **500** | 79.60- 0.12 | 79.60- 8.23 | 79.00- 8.78 | 80.00- 0.41 |
| **1000** | 48.10- 0.05 | 48.10- 3.79 | 46.70- 3.55 | 49.70- 0.19 |

(b) Dynamic scenario

| NoD | MHA | MIRA | RRATE | TEARD |
|---|---|---|---|---|
| **100** | 100- 0.21 | 100- 8.38 | 100- 10.21 | 100- 0.44 |
| **1000** | 62.70- 0.11 | 62.60- 7.22 | 62.90- 5.34 | 64.10- 0.38 |
| **2000** | 42.70- 0.10 | 42.30- 6.74 | 42.95- 3.84 | 43.75 - 0.37 |

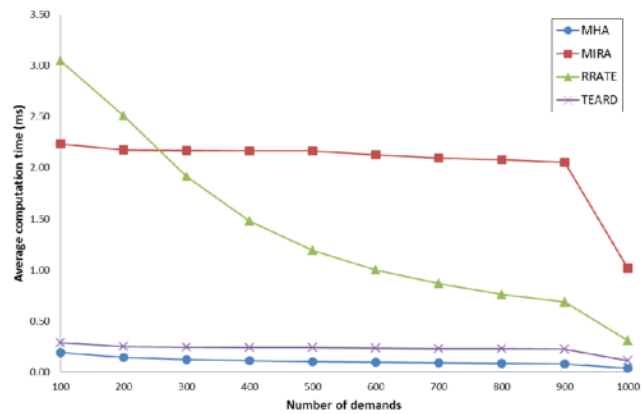

(a) Acceptance ratio

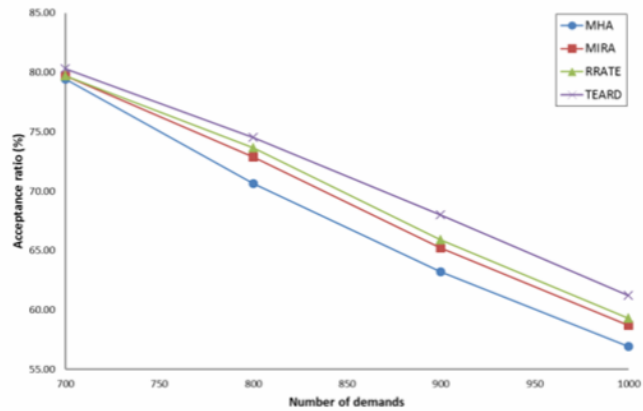

(b) Average computing time

Figure 2.  Comparison of  results of a static scenario on MIRA topo

On CESNET, TEARD has the best acceptance ratio although the differences are not noticeable as results on MIRA topology. Figure 3(a) is an example of dynamic scenario in which acceptance of TEARD is 0.8% higher than RRATE (the second best algorithm). Furthermore, MHA is better than MIRA (42.7% compares to 42.3%). The reason of good performance of MHA is that CESNET topology is inherited from a real network [9] and is optimized for the shortest path algorithm. Consequently, the improvements of TEARD over MHA on CESNET are meaningful despite small increases in the metric values. On the other hand, the large network considerably increases the computing time of MIRA and RRATE. For instance, TEARD is 10 times and 18 times faster than RRATE and MIRA respectively (figure 3(b)).

TEARD also performs impressively on ANSNET topology. Table 6 shows acceptance ratios and average computing time when the probabilities of ingress egress pairs are set 5% and 15\%; dynamic Poisson arrival time and Exponential holding time are respectively 60 and 20. Comparison in figure 4 illustrates that TEARD achieves high acceptance ratio with low computing time. Specifically, TEARD accepts 3.25% more demands and 17 times faster than RRATE. Furthermore, TEARD has compatible computing time to MHA (0.41 ms and 0.06 ms) whereas acceptance ratios is noticeably improved (47.55% and 44.65%).

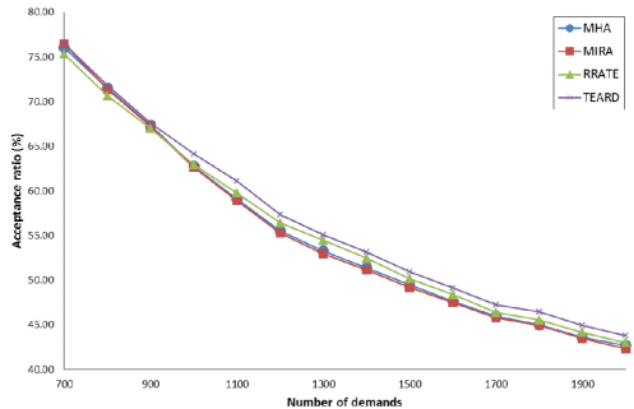

(a) Acceptance ratio

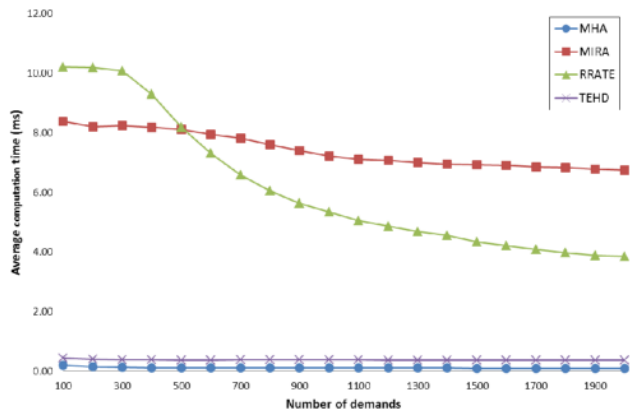

(b) Average computing time

Figure 3.  Comparison of results of a dynamic scenario on CESNET topo

Table 6.  Experiment results on ANSNET topology

(a) Static scenario

| NoR | MHA | MIRA | RRATE | TEARD |
|------|------|------|------|------|
| **100** | 100- 0.13 | 100- 13.97 | 100- 16.49 | 100- 0.47 |
| **500** | 90.80- 0.09 | 93.20- 13.06 | 86.60- 15.68 | 95.60- 0.45 |
| **1000** | 45.50- 0.03 | 46.80- 5.47 | 43.30- 6.09 | 49.90- 0.22 |

(b) Dynamic scenario

| NoR | MHA | MIRA | RRATE | TEARD |
|------|------|------|------|------|
| **100** | 100- 0.13 | 100- 13.66 | 100- 15.61 | 100- 0.47 |
| **1000** | 62.20- 0.07 | 64.10- 9.81 | 61.10- 10.99 | 67.40- 0.42 |
| **2000** | 44.65- 0.06 | 45.85- 9.16 | 44.30- 6.99 | 47.55- 0.41 |

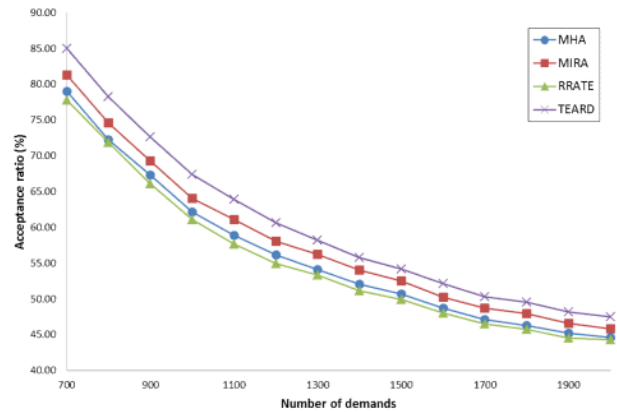

(a)  Acceptance ratio

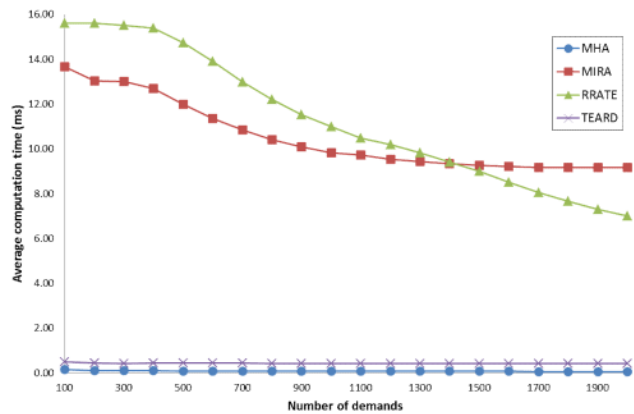

(b)  Average computing time

Figure 4.  Comparison of  results of a dynamic scenario on ANSNET topo

### 4.3. Effect of the Parameters $k_1$, $k_2$, and $k_3$

This subsection investigates how the moderation parameters affect performance of TEARD. When changing $k_i$, link weights are changed that leads to different acceptance ratios, but the computation time is not affected. 36 sets of $k_1$, $k_2$, and $k_3$ values, for each trial, are generated on the condition $0 < k_1, k_2, k_3 < 1$ and $k_1 + k_2 + k_3 = 1$. Table 7 shows the 3 highest (top 3 rows) and the 3 lowest acceptance ratios (bottom 3 rows) of the experiments presented in subsection 4.2.

As observed from table 7, variation of the moderation parameters does not significantly increase or decrease the acceptance ratio. Moreover, in many experiments, even the worst result of TEARD is better than that of other algorithms. The bottom row of table 7(b) is an example in which acceptance rate of TEARD (49.4%) is higher than MHA, MIRA (48.1%) and RRATE (46.7%) (results in table 5(a)). As a consequence, moderation values should be approximately equal (e.g. $k_1 = 0.3$, $k_2 = 0.4$, $k_3 = 0.3$ in subsection 4.2) so that all three critical parts have impacts to link weights.

Table 7. Variation of the acceptance ratio with variation of $k_1$, $k_2$, $k_3$

(a) Results of the dynamic scenario on MIRA network

| $k_1$ | $k_2$ | $k_3$ | % |
|---|---|---|---|
| 0.1 | 0.3 | 0.6 | 67.20 |
| 0.3 | 0.4 | 0.3 | 67.10 |
| 0.5 | 0.2 | 0.3 | 67.10 |
| ... | | | |
| 0.1 | 0.8 | 0.1 | 65.55 |
| 0.1 | 0.7 | 0.2 | 65.45 |
| 0.7 | 0.1 | 0.2 | 65.45 |

(b) Results of the static scenario on CESNET network

| $k_1$ | $k_2$ | $k_3$ | % |
|---|---|---|---|
| 0.2 | 0.2 | 0.6 | 50.10 |
| 0.3 | 0.2 | 0.5 | 50.00 |
| 0.4 | 0.3 | 0.3 | 50.00 |
| ... | | | |
| 0.2 | 0.1 | 0.7 | 49.60 |
| 0.4 | 0.1 | 0.5 | 49.50 |
| 0.6 | 0.2 | 0.2 | 49.40 |

## 5. CONCLUSIONS

This paper introduces TEARD, a reactive routing algorithm with minimum bandwidth constraint. The objective of traffic engineering routing is to maximize acceptance ratio so algorithms should be adaptive to routing demands. Therefore, TEARD considers not only network information but also data from historical routes. Evaluation has been conducted with different routing demands as well as different network topologies and routing scenarios. Experimental results demonstrate that TEARD accepts more demands in less computation time than other popular TE routing algorithms.

However performance of the proposed algorithm is decreased when network is large and operates in a long time because routing data becomes large. A possible solution is to limit routing data in a period of operating time instead of using the whole history data. Furthermore, if network topology changes frequently, TEARD as well as other two-phase routing algorithms are negatively affected by the time consuming offline calculation.

## AUTHORS

**Cao Thai Phuong Thanh** was born in Vietnam in 1983.
He received B.E. in Information Technology from Ho Chi Minh City University of Science, Vietnam in 2005; MSc. in Computer Science from University of East Anglia, UK in 2007, and currently a PhD candidate in Information System from Post & Telecommunications Institute of Technology, Vietnam. He is working as lecturer in Faculty of Information Technology, Saigon University, Vietnam.

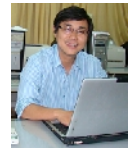

**Ha Hai Nam** was born in Vietnam in 1975.
He received B.E. in Telecommunication Engineering from University of Transport and Communications, Vietnam; MSc. in Information Technology from Hanoi University of Science and Technology; PhD. in Computer Science from Newcastle University, UK. He is, currently, lecturer of Faculty of Information Technology I, Post & Telecommunications Institute of Technology, Vietnam. His research interest includes optimization, intelligent systems, HCI, service oriented architect.

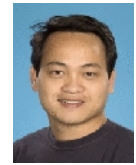

**Tran Cong Hung** was born in Vietnam in 1961.
He received the B.E in electronic and Telecommunication engineering with first class honours from Ho Chi Minh City University of Technology in Vietnam, 1987.He received the B.E in Informatics and Computer Engineering from Ho Chi Minh City University of Technology in Vietnam, 1995. He received the Master of Engineering degree in Telecommunications Engineering from Hanoi University of Technology in Vietnam, 1998. He received Ph.D at Hanoi University of Technology in Vietnam, 2004. His main research areas are B – ISDN performance parameters and measuring methods, QoS in high speed networks, MPLS. He is, currently, Associate Professor PhD. of Faculty of Information Technology II, Posts and Telecoms Institute of Technology in Ho Chi Minh City, Vietnam.

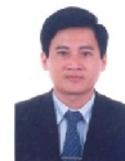